# Design Challenges for the Implementation of Smart Homes

**Nesreen Mufid, EiZ Engineering, Amman, Jordan. Email: nesreen@eiz-eng.com.jo**

## 1. Introduction

Home automation for many years had faced challenges that limit its spreading around the world. These challenges caused by the high cost of Own such a home, inflexibility system (cannot be monitored outside the home) and issues to achieve optimal security. Our main objective is to design and implement a smart home model that is simple, affordable to the users. The proposed system provide flexibility to monitor the home, using the reliable cellular network. The user will be able what is inside the home when he /she is away from home. In addition to that, our model overcome the issue of the security by providing different sensors that detects smoke, gas , leakage of water and incases of burglary [1]. Moreover, a camera will be available in the home to give a full view for the user when he /she is outside the home. The user will be informed by an application on his/she phone incase if there is a fire ,water leakage and if someone break into the house . This will give the user a chance to take an action if such cases happened. Furthermore, the user can monitor the lighting system of the home, by giving the user a chance to turn the lights on and off remotely [2].

## 2. Design criteria

The proposed Model include various features to provide an intelligent and safe environment inside the home for the home members when they are away from it. In order to build such a system, the selection of different electrical devices and components in the system should match our criteria [3]. After a deep studying of the previous projects of smart homes, we have chosen our design criteria based on the challenges discussed previously. These criteria of design are:

1. Cost [4]
2. Size [5]
3. Reliability [6]
4. Complexity [7]
5. Availability [8]

**Cost**: One the challenges that encountered by previous smart home projects is the high cost of the system. In order to attract the people and convenience them to buy such product, the cost should be affordable to average people. In this design, we are targeting to attract people by providing an affordable design yet effective that accomplish our vision about a smart home [9].

**Size:** In today's technology, we are interested to buy the newest and modern devices that is easy to carry at the same time. For instance, smartphones, tablets … etc. In our design for the smart home [10], we are looking for smaller electrical devices (e.g sensors, UMTS and microcontroller [11]) to build a prototype of a smart home. In addition to that, physical demission of the components plays an important role in such mobile communication devices, especially when changing and repairing those devices [12].

**Reliability:** is one of the critical factors that determines if the system is operating successfully or not. In addition to that, in M2M communication standards such as Bluetooth or Zigbee or cellular network reliability [13] plays an important role for selection of what communication standard that suits the application .In our design of a smart home, we are giving the user a chance of higher reliability of monitoring the home appliances. Because the automation system inside the home is connected over a cellular network that gives more mobility and reliability to the system [14].

**Complexity**: The system should be simple to install and address by the user. It should function smoothly and not to add more complexity to the home. In addition to that, the system should be simple yet effective to attract the communications company to adopt the concept [15].

**Availability**: The components that exist in the smart home should be available in the market. This will help the user to repair or install different components inside the home. The more availability the components are the better chance to find it in a cheaper price [16].

## 2.1. Microcontroller

In order to control basic home function remotely and before going through the installation of the desired sensors, A Microcontroller (tiny computer) should be placed in between so that it can control several digital and analog inputs, create programs that can interact with the physical world, translate inputs' values into written information that the user can understand and set certain thresholds. The term microcontroller is broadly used; the construction of a microcontroller is equivalent to a simple computer located in a single chip. It is programmed to do some simple task like reading input from sensors or starting a small DC motor [17].

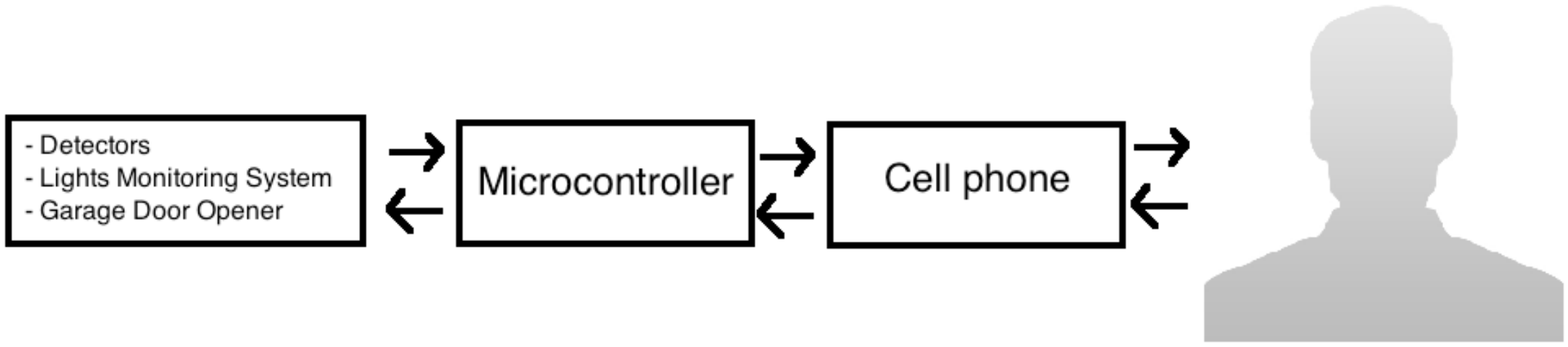

*Figure 1*: *Block diagram of the smart home design*

### 2.1.1. Why Arduino?

An integrated circuit that might include millions of logic gates that can be electrically configured to perform certain task is called a Field Programmable Gate Array or FPGA. It can be reprogrammed to do several logic tasks by operating the task into a certain number of logic gates. FPGA by itself can do nothing; it depends upon the users' digital circuit and configuration file. The main reason that makes FPGA flexible, the user can configure it as many times as he wants. However, the price of this flexibility is already paid by its high power consumption compared to the typical microcontroller [18]. Moreover, FPGA would consume longer time to define the whole code from the beginning and the converting it to machine language. To manufacturers, the price of FPGA is considered expensive in comparison with simple microcontrollers, but they still use it when it comes to high complexity with low demand. The process of configuring an FPGA starts with writing some HDL, either by using Verilog or VHDL. After that, the HDL should be synthesized into a bit file in order configure the FPGA. This configuration must be done each time the FPGA is powered on, because it stores the configuration in RAM [19].

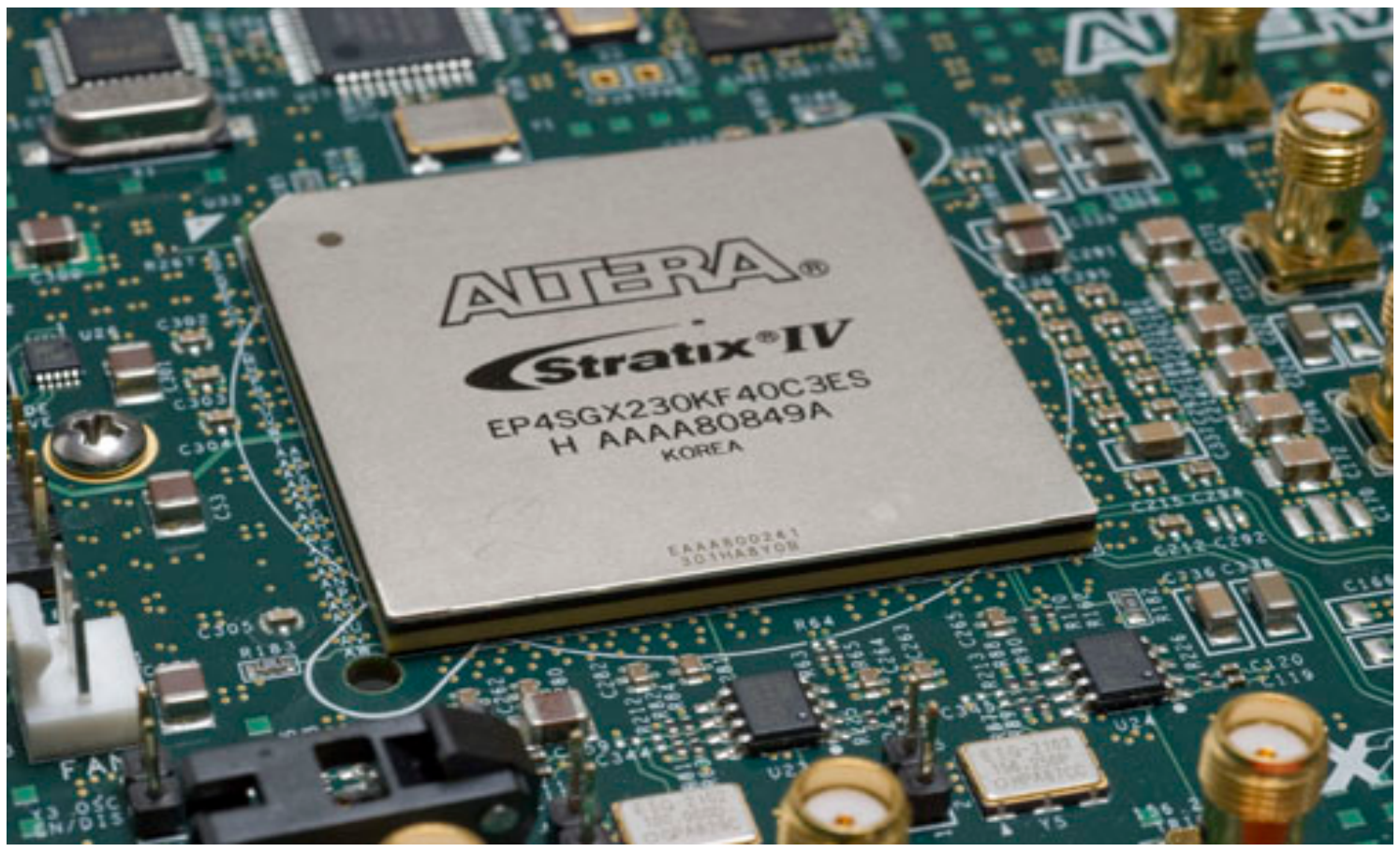

*Figure 2*: *FPGA chip*

Arduino is an open source hardware and software platform based on flexibility and simplicity. In the last few years, designers around the world trusted this simple microcontroller to create their interactive objects and projects. It is truly related to the human environment by receiving inputs from many types of sensors and delivering output to control lights, motors and actuators. IDE (a graphics programming language and development system popular with artists and designers) is used to program Arduino; it works on Windows, Macintosh and Linux. It has a fast compiler and thousands of libraries to help the programmer to do complex things like communicating with LCD screens, GPS, SD cards, servomotors and many other useful appliance using simple coding shortcuts [20]. The attached analog-to-digital input makes this microcontroller cover 99% of the low cost sensors on the market interface easily with Arduino. The price of Arduino starts from 30$ and it is expected to reach 20$ soon.

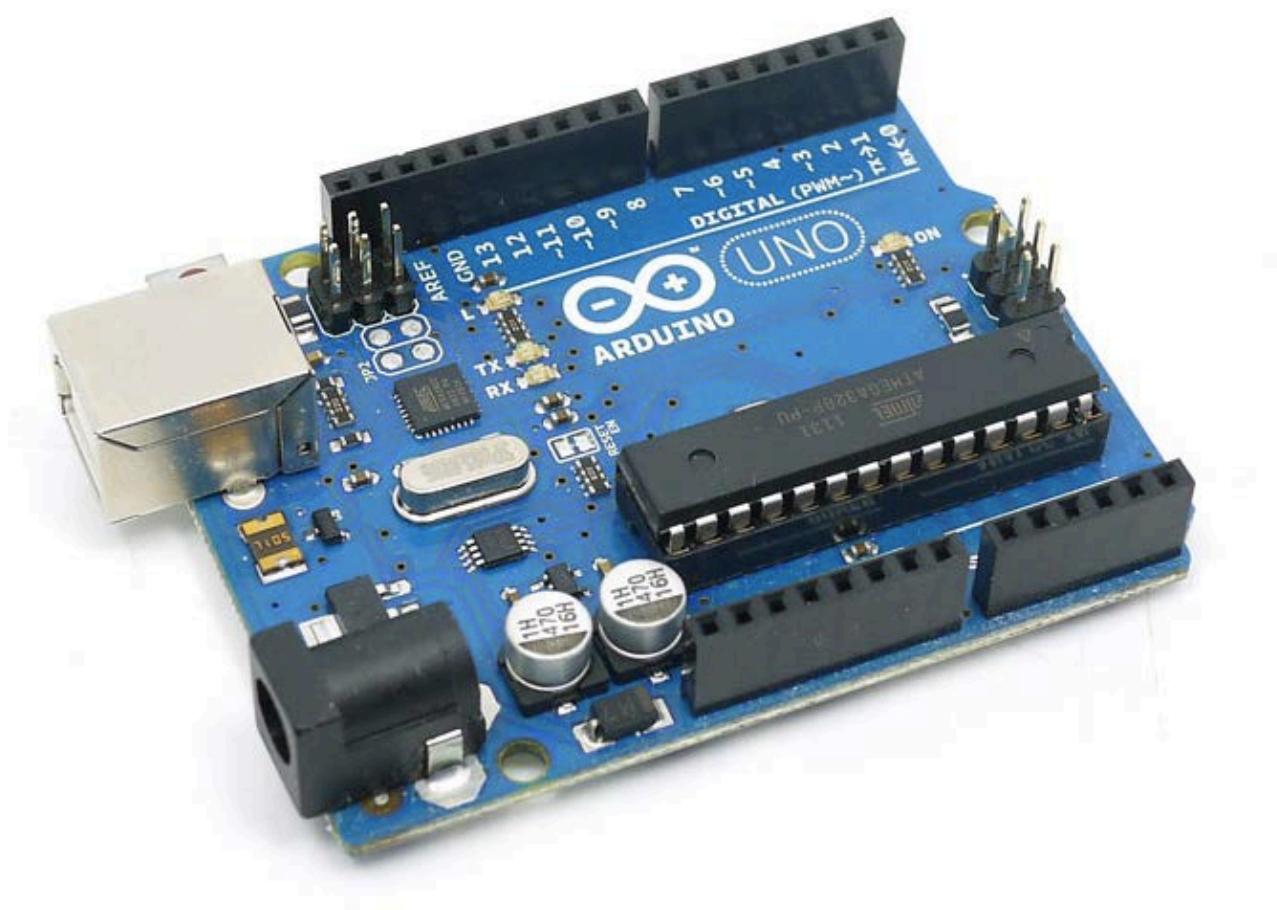

*Figure 3*: *Arduino UNO*

In summary, the advantages stated above are clearly and strongly supports the selection criteria of this project. Arduinos consumes less power than FPGAs. FPGAs take longer time in configuration while Arduinos has a high uploading speed. Building chips with FPGAs costs more than Arduinos. Arduino could be used to build a great software with multiple feature by using the open source in short period of time, while with FPGA, the same high quality could be driven but after spending time that increases proportionally with the complexity issues that will rise frequently.

## 2.2. Temperature sensor

To measure temperature, there are many special types available. The most three common are resistance temperature detectors (RTDs), Thermistors, Thermocouples and Integrated Circuit (IC) temperature sensors. Analog IC solid-state sensors are the most suitable solution for this project because they provide their output as current or voltage proportionally with temperature without need for extra circuitry. In addition, IC sensors considered being very competitive in cost with the other temperature sensing technologies, in some situations less pricy than RTD's and Thermistors.

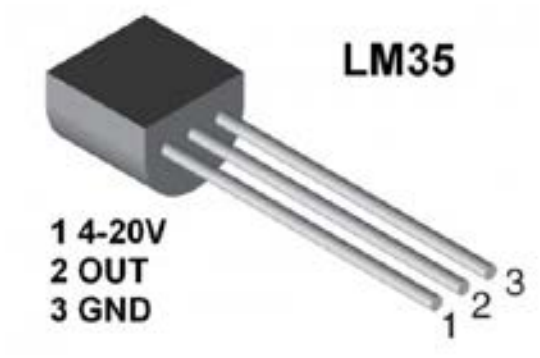

*Figure 4: LM35 Temperature sensor*

Going deep into IC's area, a comparison between three types of IC sensors was made and the selection criteria was based on the accuracy and the price as shown in table 2.1.

*Table 1: Temperature ICs comparison*

| | LM35 | TC74A | ADT7420 |
|---|---|---|---|
| Output | Analog | Digital | Digital |
| Accuracy | ±0.5 ℃ | ±3 ℃ | ±0.25 ℃ |
| Price | $1.05 | $1.62 | $5.46 |
| Operating Range | 4 – 30 V | 3 – 5.5 V | 2.7 – 3.3 V |

The popular LM35 temperature sensor creates an analog voltage directly proportional to temperature; the scale is explained in the datasheet that each 0.1℃ produces an output of 1 millivolt (10mV per degree)[Appendix A]. The following code converts the analog read values into millivolts and then divides these values by 10 to get it in degree. Also an LED is connected to pin 13 in order to trigger it as an alarm when the temperature of the smart home is above 25℃. The sensor accuracy is about 0.5℃, which is why integer math is used instead of floating point.

## 2.3. Smoke and gas leakage detector

Gas leakage and smoke are one of the most concerns that could make the homeowner worried when he is away from home. Installing gas leakage and smoke detectors in the smart home can participate effectively in achieving the security-monitoring objective.

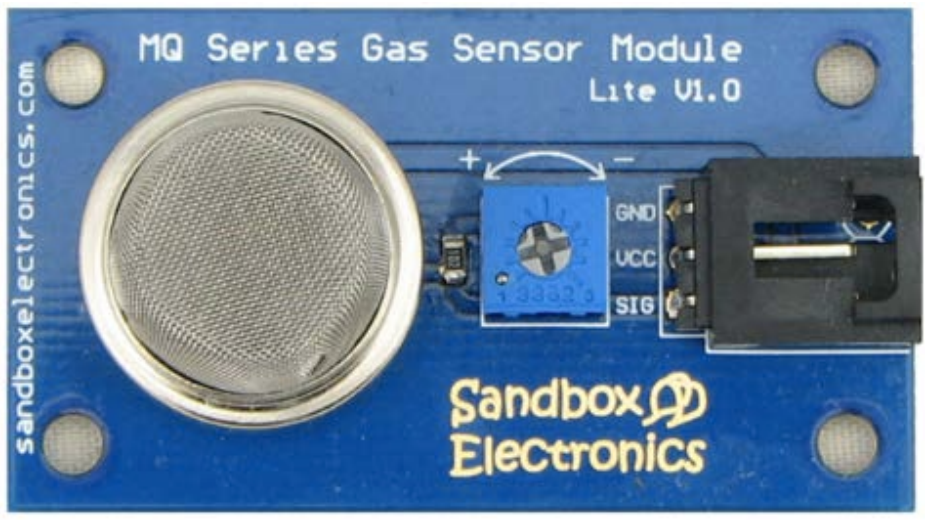

*Figure 5: MQ gas sensor module*

The MQ-2 sensor module has been utilized as the sensitive component; the module board has an adjustable resistor and a protection resistor. It can be used to sense propane, alcohol, hydrogen, LPG and smoke. The resistance changes as the concentration of the desired gas changes. In addition to the low cost of the MQ-2 gas sensor module, also the compact size was the second reason for selecting this module [21].

The adjustable (0 - 50 $k\Omega$) and the protection (4.7 $k\Omega$) resistors are in series, which forms a load resistor (4.7 – 54.7 $k\Omega$). The output of the voltage divider that was formed by the sensor's resistance $R_s$ and $R_L$ could be read by the Arduino. Calibration is most important stage in getting an accurate reading from gas sensor. Resistor named $R_0$ is calculated during calibration by sampling reading when the sensor is in clean air [22]. After that, the concentration of the smoke and LPG could be calculated by dividing $R_s$ over $R_L$ and use this ratio as an input (code in appendix B).

## 2.4. Motion Detector

Currently, the best solution to overcome house-burgling problem is to place motion detectors that increases the home security when homeowner is not in house. Motion detection is usually aiming on developing the home security monitoring with motion sensor controlled by a microcontroller. More motion sensor should be placed in order to reach the highest level of home security monitoring [23]. Motion detection is the process triggering an alarm after identifying physical movement in a specified area. The motion sensors are usually placed in main doorways or windows. Two popular motion sensors are the ultrasonic and the passive infrared (PIR); the ultrasonic type is frequently used for automatics door openers for the reason of its high accuracy in measuring how far is the detected object from the sensor. On the other hand, the PIR is preferred to be used in the area where detecting if there is person in the room is the desired objective. It works by detecting the heat energy of a moving object such as human body [24].

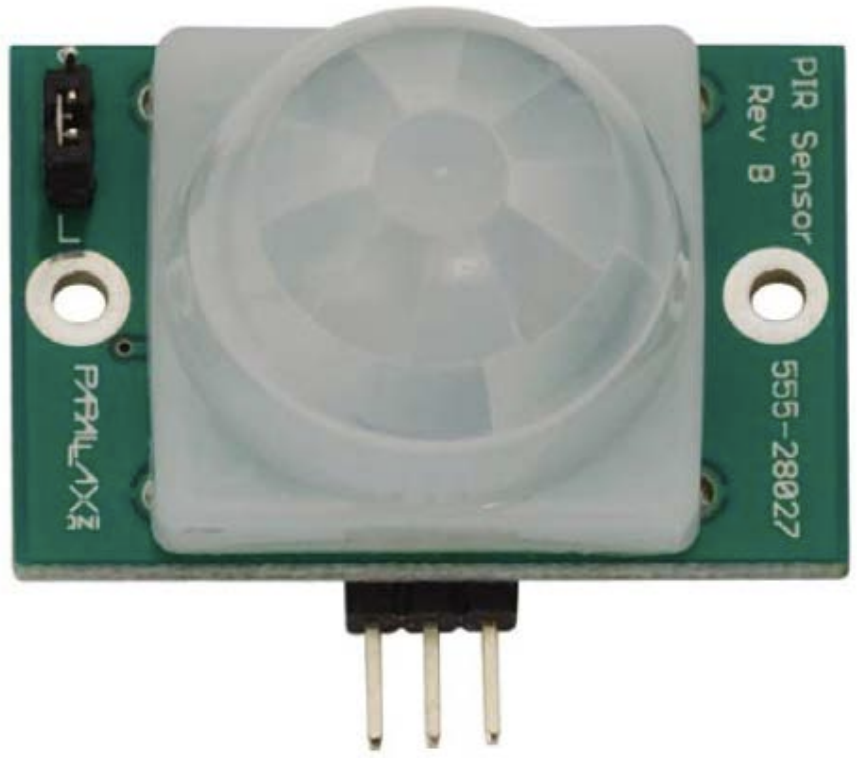

*Figure 6: PIR motion sensor*

The PIR simplicity in changing digital values when someone moves nearby makes it the suitable solution to connect to the Arduino. The following code represents an algorithm to read if the sensors' output is high or low, because the PIR is acting like a pushbutton that gives high value when motion is detected and low value when there is no motion.

## 2.5. Water Leakage Detector

Assuming the dot that stands at the end of this sentence was representing a hole in home's water system. At the first look it might not seem worth tracking, but it can waste more than 15,000 Liters by its size only – enough for showering once daily through a year period. Water leakage detection is very important for energy saving when water issues rises up, because it had been a major problem for many homeowners [25]. The problem increases for those who had to stay outside home for hours or go for vacations frequently. They will have no worries about their water bill if they ensured that their home's water system is safe and secured. However, traditional leakage detection systems such as geophones and radar technique are expensive and time consuming. On the other hand, Remote leakage detection is considered time and cost effective compared with the traditional ones. The idea of this part of the smart home project is to let the Arduino send SMS to the homeowner telling him that there is a leakage when water leakage is detected [26]. After that, the user can take action either by telling someone nearby home to close main valve quickly or by a better solution depending on his location at the leakage time.

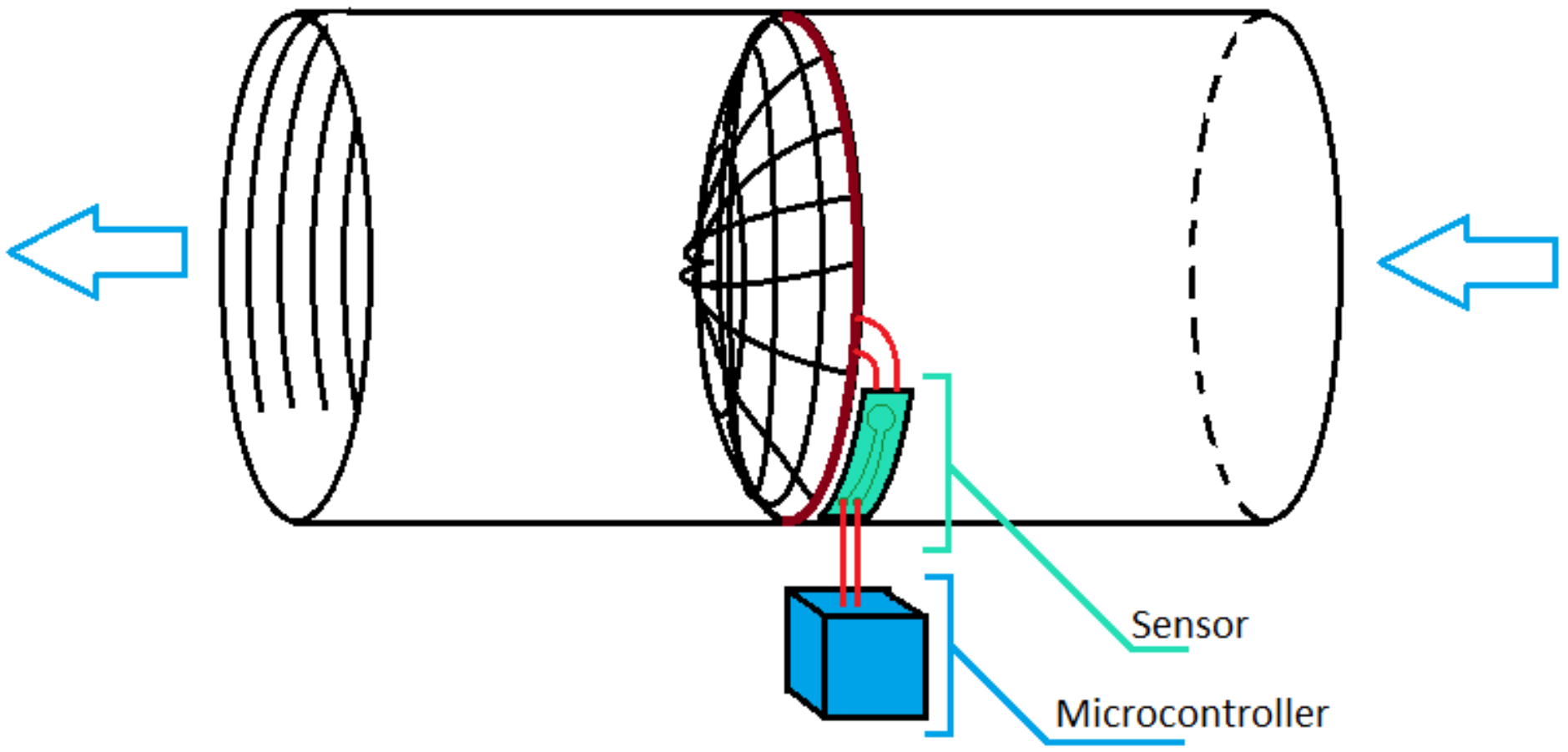

*Figure 7: Water leakage Detection prototype(Designed by EE student in Qatar University)*

As shown in fig. 5, a piece of thin rubber that has wide hole with a sieve shape will be placed inside the pipe of the main home tank. It should be placed in the main pipe in order avoid placing detector in each pipe. High sensitivity vibration sensor is attached to the rubber. Water flow will make the rubber. Sieve vibrate and the vibration sensor will detect this vibration indicating that there is a leakage somewhere at home [27]. In order to detect unnoticeable leakages such, it is effective to connect Op-Amp amplifier to the output of the vibration senor so the Arduino can read the resultant microvolts caused by the leakage of the small water drops.

## 2.6. Home Lights Monitoring System

One of the main of feature of a smart home is the ability to manage and the energy consumption by controlling the lights systems. In this design of smart home, the user will be able to monitor the lights inside the home remotely when he /she are away from it [28]. Sometime the lights are forgotten to be turned off when the home members leave the home .By using the phone the user can send a command to turn the lights on and off remotely without being inside the home.

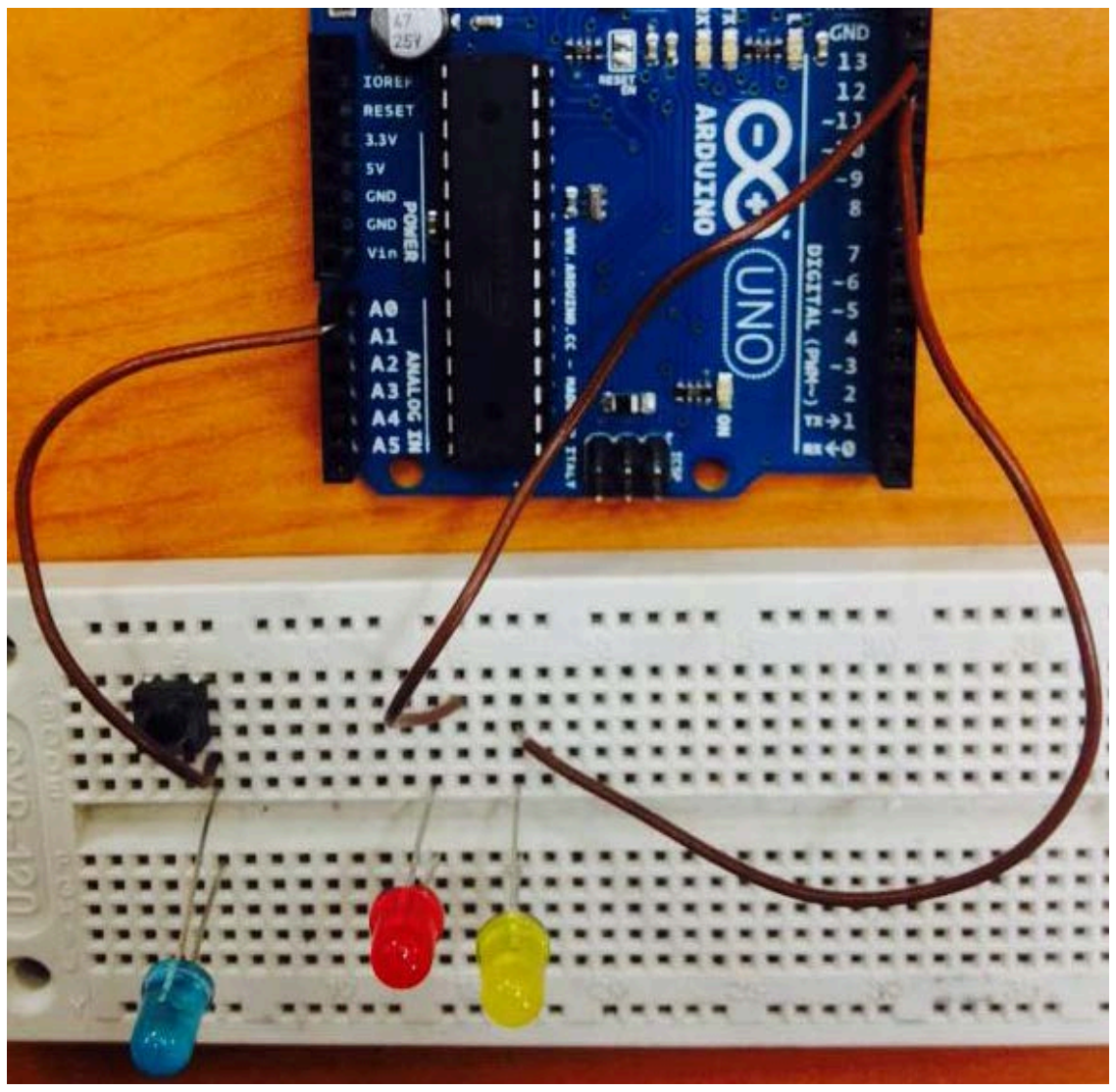

*Figure 8: Connections of the light monitoring system protoype*

Lights Monitoring system prototype was designed using three LEDs, switch and an Arduino as well. Analog pin A0 is connected to the node between the switch and the blue LED that represents the room light. The idea is that when light switch is turned on, the microcontroller will trigger pin 13 (Red LED) indicating that the user forgot to turn off home lights and will send SMS alerting the user. Rather than that, the Yellow LED will remain on to indicating that the home lights are off. The figure shows the code used to do this task.

## References

[1] E. Datsika et.al., “Software Defined Network Service Chaining for OTT Service Providers in 5G Networks”, IEEE Communications Magazine, 55 (11), 124-131, November 2017.

[2] S. Alanazi et.al., “Reducing Data Center Energy Consumption Through Peak Shaving and Locked-in Energy Avoidance”, IEEE Transactions on Green Communications and Networking 1 (4), 551-562, August 2017.

[3] Michael Margolis, Arduino Cookbook, 2nd ed., Shawn Wallace and Brian Jepson, Eds. Newyork, USA: O'Reilly Media, Inc, 2012.

[4] E. Kartsakli et.al., “A Threshold-Selective Multiuser Downlink MAC scheme for 802.11n Wireless Networks”, IEEE Transactions on Wireless Communications, April 2011.

[5] R. Imran et.al., “Quality of Experience for Spatial Cognitive Systems within Multiple Antenna Scenarios”, IEEE Transactions on Wireless Communications, vol. 12, no.8, August 2013.

[6] Vini Madan and S.R.N, "GSM-Bluetooth based Remote Monitoring and Control System with Automatic Light Controller ," International Journal of Computer Applications , vol. 46, no. 0975 - 8887, May 2012.

[7] A. Elbery et.al., “IoT-Based Crowd Management Framework for Departure Control and Navigation”, IEEE Transactions on Vehicular Technology 70 (1), 95-106, January 2021.

[8] B. Hamdaoui et.al., “IoTShare: A Blockchain-Enabled IoT Resource Sharing On-Demand Protocol for Smart City Situation-Awareness Applications”, IEEE IoT Journal 7 (10), 10548-10561, November 2020.

[9] Dale Cigoy, "How to Selcet the Right Temperature," Keithyley Instruments, Inc., Cleveland, Electrical 2816, 2007.

[10] B. Hamdaoui et.al., “Dynamic Spectrum Sharing in the Age of Millimeter Wave Spectrum Access”, IEEE Network 34 (5), 164-170, July 2020.

[11] A. El-Wakeel et.al., “Robust Positioning for Road Information Services in Challenging Environments”, IEEE Sensors Journal 20 (6), 3182-3195, December 2019.

[12] Toril Laberg, Haakon Aspelund, and Hilde Thygesen, "SMART HOME TECHNOLOGY, Planning and management in municipal services," Directorate for Social and Health Affairs, the Delta Centre, Oslo, 2005.

[13] N. Zorba and A.I. Pérez-Neira, “Opportunistic Grassmannian Beamforming for Multiuser and Multiantenna Downlink Communications”, IEEE Transactions on Wireless Communications, no. 4, April 2008.

[14] Ali Isilak, "SMART HOME APPLICATIONS FOR DISABLED PEOPLE BY USING WIRELESS SENSOR NETWORK ," Department of Computer Engineering, Yeditepe University, Faculty of Engineering and Architecture , Engineering Project 2010.

[15] N. Zorba and A.I. Pérez-Neira, "Robust Power Allocation Schemes for Multibeam Opportunistic Transmission Strategies Under Quality of Service Constraints", IEEE JSAC special issue on MIMO for Next-Generation Wireless Networks, no.8, August 2008.

[16] Adiline Macriga, "Smart Home Monitoring and Controlling System Using Android Phone," International Journal of Emerging Technology and Advanced Engineering, vol. 3, no. 11, pp. 426,427, November 2013.

[17] E. Datsika et.al., "Cross-Network Performance Analysis of Network Coding Aided Cooperative Outband D2D Communications", IEEE Transactions on Wireless Communications vol. 16, May 2017.

[18] Alan G. Smith, Introduction to Arduino., 2011.

[19] B. Khalfi et.al., "Efficient Spectrum Availability Information Recovery for Wideband DSA Networks: A Weighted Compressive Sampling Approach", IEEE Transactions on Wireless Communications, vol. 17, April 2018.

[20] Meensika Sripan, Xuanxia Lin, Ponchan Petchlorlean, and Mahasak Ketcham, "Research and Thinking of Smart Home Technology," in International Conference on Systems and Electronic Engineering, Thailand, 2012.

[21] Z. Wu et.al., "Device-to-Device Communications at the TeraHertz band: Open Challenges for Realistic Implementation", IEEE Communications Standards Magazine, 7 (1), 82-87, December 2022.

[22] A.Z. Abyaneh et.al., "Empowering Next-Generation IoT WLANs Through Blockchain and 802.11 ax Technologies", IEEE Transactions on Intelligent Transportation Systems, August 2022.

[23] Adiline Macriga, "Smart Home Monitoring and Controlling System Using Android Phone," International Journal of Emerging Technology and Advanced Engineering, vol. 3, no. 11, pp. 426,427, November 2013.

[24] M. Alzard et.al., "Performance Analysis of Resource Allocation in THz-based Subcarrier Index Modulation Systems for Mobile Users", IEEE Access 9, 129771-129781, September 2021.

[25] A. El-Wakeel et.al., "Towards a Practical Crowdsensing System for Road Surface Conditions Monitoring", IEEE Internet of Things Journal 5 (6), 4672-4685, March 2018.

[26] ULTRAN electronic components, "LISA-U2 series 3.75G HSPA+ Wireless Modules," Datasheet, 2023.

[27] B. Khalfi et.al., "Optimizing Joint Data and Power Transfer in Energy Harvesting Multiuser Wireless Networks", IEEE Transactions on Vehicular Technology 66 (12), 10989-11000, June 2017.

[28] M. Aboualola et.al., "Edge Technologies for Disaster Management: A Survey of Social Media and Artificial Intelligence Integration", IEEE Access, vol. 9, August 2023.